\documentclass[12pt]{article}
\usepackage{amsmath}
\usepackage{mathbbol}
\begin{document}
\renewcommand{\thefootnote}{\fnsymbol{footnote}}
\newcommand{\e}{{\bf E}}
\newcommand{\m}{{\bf B}}
\newcommand{\h}{{\bf H}}
\renewcommand{\v}{{\boldsymbol \varPhi}}
\newcommand{\V}{\Phi}
\newcommand{\D}{\Delta}
\renewcommand{\l}{{\cal L}}
\renewcommand{\c}{\cdot}
\newcommand{\p}{\partial}
\renewcommand{\P}{{\bf\Pi}}
\newcommand{\n}{\boldsymbol{\nabla}\V}
\newcommand{\G}{\Gamma}
\renewcommand{\d}{\boldsymbol{\nabla}}
\renewcommand{\b}{\boldsymbol{\G}}

\rightline{}

\vspace{0.6cm}

\begin{center}
{\Large {\bf D3-brane intersecting with dyonic BIon}}

\vspace{1truecm}

\large 
S. Tamaryan,${}^{a,b}$\footnote[1]
{E-mail:sayat@moon.yerphi.am}
H.J.W. M\"uller-Kirsten${}^b$\footnote[2]
{E-mail:mueller1@physik.uni-kl.de}
and D. K. Park${}^c$\footnote[3]
{E-mail:dkpark@hep.kyungnam.ac.kr}

\vspace{0.8cm}

\normalsize
${}^a${\it  Theory Department, Yerevan Physics Institute,\\ 
Yerevan-36, 375036, Armenia}\\
\vspace{0.4cm}
${}^b${\it Department of  Physics, University of Kaiserslautern,\\ 
67653 Kaiserslautern, Germany}\\
\vspace{0.4cm}
${}^c${\it Department of Physics, Kyungnam University,\\ 
Masan, 631-701, Korea}\\
\end{center}

\vspace{0.6cm}

{\centerline {\bf Abstract}}

\vspace{0.3cm}

The Born--Infeld theory of a D3-brane intersecting with
$(p,q)$ strings is reconsidered. From the assumption that the 
electromagnetic fields are those of a dyon, and using
 the kappa invariance of the action, the explicit scalar field and its charge
are derived. Considering perturbations orthogonal to both branes, the 
$SL(2,\mathbb{Z})$-invariant S-matrix is obtained. Owing to the selfduality 
of the brane 
the latter can be evaluated explicitly in both high and low energy regions.

\thispagestyle{empty}

\newpage

\section{Introduction}

\noindent

The Dp-brane Born--Infeld action has BPS-saturating solutions which have  a 
spacetime interpretation as intersections with other branes \cite{cm, gib}.
These 1/2 supersymmetric solutions appear as spikes on the world--volume with 
uniform energy per unit length and their 
infinite energy is associated with an 
infinite string (or brane) of fixed tension. The sources of these BIonic 
solutions on a brane appear as the endpoints of  strings
 outside the 
brane. The fundamental string attached to the
brane is described by the  electric point 
charge solution. In the case of the D3-brane there are dyonic BIons 
corresponding  to  $(p,q)$ strings
of infinite length  that end on the D3-brane.

The perturbations propagating via
the  attached fundamental string reveal specific 
nonlinearities of the BI action. 
The equation describing these fluctuations
 was shown to be 
convertible into a modified Mathieu equation \cite{gubh, mmlz, ptmz}
which allows  an   explicit and exact  expression  for the S-matrix 
to be obtained  as a function of the 
frequency of the fluctuations.
In the following we generalise these considerations
to  the case of the full $(p,q)$ strings
in the background  and show that
the S-matrix is a function of the   
product of the
$SL(2,{\mathbb Z})$-invariant  tension of the
 $(p,q)$ strings and the frequency of
the fluctuation. It turns  out that as the frequency  increases, the
S-matrix  goes over from perfectly reflecting to perfectly absorbing;
these considerations require a knowledge of the S-matrix in both the
low-frequency  and the high-frequency regions.
The reflection-absorption behaviour obtained
 is in full compliance with Pochinski's
 picture of D-branes as boundaries 
for open strings.

The basic procedure to obtain  such type of solutions is
to consider  the BPS limit 
\cite{hlw, ggt, gkmtz}. In this way finite length
 strings stretched between 
two D3-branes together with  the string junctions
 were obtained in \cite{gkmtz}.
 One of the  
D3-branes was considered as a 
IIB supergravity solution and taken  as a 
background for another D3-brane. Static BIon solutions of this type have a 
finite energy proportional to  the
distance between the branes  with the $(p,q)$ string tension as 
coefficient. They preserve 1/2 of the supersymmetries,
 while string junctions 
preserve 1/4 of the  supersymmetries. 
A further  interesting development was achieved 
in \cite{ht, hh}, where the 't Hooft--Polyakov monopole of Yang--Mills theory  is 
analysed in 3+1 dimensions.
 The Higgs scalar was interpreted  as a  transverse 
scalar for a two D3-brane configuration  connected by the  monopole 
solution. It was found  that part of
the  D-string flux lives on the D3-brane, while 
another part passes through the space between the branes. 

The D3-brane carries Fundamental and Dirichlet string charges \cite{evs, tub}.
In the BPS limit the brane becomes tensionless 
so that its energy is  entirely due to 
 the electromagnetic field  and is the sum of the electric and magnetic 
charges \cite{bak, hyp, vv}. Then the D3-brane can be interpreted as a 
blown up $(p,q)$ string which is the  three-dimensional analog of
 supertubes  
\cite{mt}. 

It is unclear whether the BPS bound can be
obtained from examination of the  complicated 
Hamiltonian under consideration. However, there is an alternative 
approach which we pursue  here.
In Section {\bf 2} we present the action, Hamiltonian and equations of motion.
In Section {\bf 3} we first accept that electric intensity and magnetic 
induction are those created by the  dyon and 
next find the explicit expression for 
the scalar field using the supersymmetry argument. The  solution obtained 
satisfies the equations of motion and preserves  half of the supersymmetry.
In Section {\bf 4}
the equation of small fluctuations is derived and converted into
a modified Mathieu equation. Investigation of the resulting
S-matrix then leads to the conclusions referred to above.

\section{Equations of motion}

\noindent

The D3--brane action with constant dilaton and 
vanishing Kalb--Ramond fields is
 given by
\begin{equation}
\label{bi}
I=-T_3\int d^4\xi\,\sqrt{{-\rm det}(g_{\alpha\beta}+2\pi F_{\alpha\beta})}\,\,,
\end{equation}
where $g_{\alpha\beta}$ is the induced metric and $F_{\alpha\beta}$ is the 
electromagnetic field strength. 

To construct a D3--brane intersecting with $(p,q)$ strings we keep the 
electric field $\e$, the magnetic induction $\m$
(both with the factor $2\pi$ absorbed)
 and a transverse scalar 
field $\Phi$ in the ninth direction. The freedom of the world--volume 
diffeomorphism invariance is fixed by choosing the static gauge for which the 
world--volume coordinates $x^{\alpha}\,(\alpha=0,1,2,3)$ are equated with 
the first four spacetime coordinates of the flat target space. The action  
reduces to
\begin{equation}
\label{act}
I=T_3\int\l\, d^4x\,\,,
\end{equation}
where the Lagrangian density $\l$ is
\begin{equation}
\label{lagr}
\l=-\sqrt{1-\e^2+\n^2+\m^2+
(\m\cdot\n )^2-(\e\cdot\m)^2-|\e\times\n|^2}\,\,.
\end{equation}
The conjugate momentum $\P$ associated with the electromagnetic potential is 
given by
\begin{equation}
\label{ind}
\P=\frac{\p\l}{\p\e}=-\frac{1}{\l}[\,(1+\n^2)\e+(\e\cdot\m)\m-
(\e\cdot\n )\n\, ]\,.
\end{equation}
This  nonlinear relation  can be reversed to give the electric field $\e$ in 
terms of canonical momentum $\P$, magnetic induction $\m$ and scalar field 
$\Phi$,
\begin{equation}
\label{rev} 
\e=\frac{v\,\P-w\,\m+[\,v\,(\P\c\n)
-w\,(\m\c\n)\,]\n}{\sqrt{(1+\n^2)(uv-w^2)}}\,,
\end{equation}
where the quantities $u,v,w$ are defined as
\begin{eqnarray}
\label{uvw}
u&\equiv&1+\P^2+\n^2+(\P\c\n)^2\,,\nonumber\\
v&\equiv&1+\m^2+\n^2+(\m\c\n )^2\,,\nonumber\\
w&\equiv&\P\c\m+(\P\c\n)\,(\m\c\n)\,.
\end{eqnarray}
After performing the  Legendre transform
\begin{equation}
\label{leg}
\cal H=\e\c\P-\l,
\end{equation}
the Hamiltonian density $\cal H$ appears as
\begin{equation}
\label{ham}
{\cal H}=\sqrt{\frac{uv-w^2}{1+\n^2}}\,.
\end{equation}
The Hamiltonian (\ref{ham}) is invariant under the electric--magnetic duality 
rotation through an arbitrary angle $\theta$
\begin{eqnarray}
\label{dual}
\begin{array}{c}
\P\to\P\cos\theta+\m\sin\theta\,,\\
\m\to\m\cos\theta-\P\sin\theta\,.
\end{array}
\end{eqnarray}
The invariance (\ref{dual}) takes solutions of equations of motion into 
solutions and permits to construct a set of S--dual solutions.

The equations of motion are conveniently presented by introducing the 
magnetic field strength $\h$,
\begin{equation}
\label{int}
\h=-\frac{\p\l}{\p\m}=-\frac{1}{\l}[\,\m-(\m\cdot\e)\,\e+(\m\cdot\n)\n\,]\,,
\end{equation}
and the spatial part $\v$ of the four-momentum of the scalar field 
\begin{equation}
\label{v}
\v=-\frac{\p\l}{\p\n}=-\frac{1}{\l}\,[\,(1-\e^2)\n+(\e\cdot\n)\,\e+
(\m\cdot\n)\,\m\,\,]\,.
\end{equation}
The Euler--Lagrange static equations are
\begin{subequations}
\label{eil-lagr}
\begin{equation}
\label{gaus}
\d\cdot\P=0\,,
\end{equation}
\begin{equation}
\label{elind}
\d\times\h=0\,,
\end{equation}
\begin{equation}
\label{scalar}
\d\cdot\v=0\,,
\end{equation}
\end{subequations}
supplemented by the corresponding Bianchi identities
\begin{subequations}
\label{bianc}
\begin{equation}
\label{magind}
\d\times\e=0\,,
\end{equation}
\begin{equation}
\label{mon}
\d\cdot\m=0\,.
\end{equation}
\end{subequations}

The terms containing the time derivative of the scalar 
field $\dot\Phi$ have been 
omitted in the Lagrangian (\ref{lagr}) for  simplicity. Considering that
$$\frac{\p\l}{\p\dot\Phi}\bigg|_{\dot\Phi=0}=0\,,$$
these terms give no contribution to the Hamiltonian (\ref{ham}),  nor  
to the equations of motion (\ref{eil-lagr}) in the static case.

\section{D3-brane intersecting with $\boldsymbol{(p,q)}$ string}

\noindent

By $SL(2,\mathbb{Z})$ invariance, both Fundamental and Dirichlet strings 
can end on the self--dual D3--brane, the former ending on electric charges and 
the latter on magnetic charges. The endpoint of a $(p,q)$ string will appear 
in the world--volume as a dyon. This idea   together with  
supersymmetry arguments  allows to construct a D3--brane intersecting with a 
$(p,q)$ string. 

\subsection{Supersymmetric solution}

\noindent

Pointlike electric and magnetic charges  at the 
same point are sources of 
an electromagnetic field with  spherical symmetry and components
\begin{equation}
\label{sol1}
\e=e_0\,\frac{\bf r}{r^3}\,,\qquad\m=b_0\,\frac{\bf r}{r^3}\,,
\end{equation}
where $e_0$ and $b_0$ are constants. The preserved part of the supersymmetry 
of the IIB Minkowski vacuum is defined by the condition
\begin{equation}
\label{k3}
\G\varepsilon=\varepsilon\,,
\end{equation}
where  $\G$ is the projection operator appearing in the $\kappa$-symmetry 
transformations. In the case under consideration it is given by \cite{ced, bk}
\begin{equation}
\label{k1}
\G=\frac{1}{\l}(\G_I\otimes I-\G_J\otimes J)\,,
\end{equation}
where
\begin{eqnarray}
\label{k2}
&\b\!\!\!&=(\G_1\,,\G_2\,,\G_3)\,,\\
&\G_I\!\!\!\!&=\G_{0123}-(\n\c\b)\G_{01239}+\e\c\m\,,\\
&\G_J\!\!\!\!&=(\e\c\b)\G_{123}-((\e\!\times\!\n)\c\b)\G_9+
(\m\c\n)\G_{09}-(\m\c\b)\G_0\,,
\end{eqnarray}
and
\begin{equation}
\label{sig}
I=i\sigma_2\,,\quad J=\sigma_1\,.
\end{equation}
The 10 dimensional matrices
 $\G_{\alpha}$ act on the Weyl--Majorana index while 
the matrices $I,J$ act on the $SO(2)$ index of the spinor. The matrices 
$\G_I,\G_J$ have the properties
\begin{equation}
\label{com}
[\G_I\,,\,\G_J]=0\,,\quad\G_J^2-\G_I^2=\l^2\,\mathbb{1}\,.
\end{equation}
We denote the $SO(2)$ components of 
the spinor $\varepsilon$ by $\varepsilon_1$ 
and $\varepsilon_2$ and set
\begin{equation}
\label{k4}
\varepsilon_1=\G_{123}\epsilon, \,\quad\varepsilon_2=-\G_0\epsilon\,.
\end{equation} 
Equation (\ref{k3}), omitting the vanishing term $\e\times\m$, splits into 
two equations 
\begin{subequations}
\label{k5}
\begin{equation}
\label{k5a}
-\frac{1}{\l}[(1+\m^2)\G_{123}-(\m-\G_{123}\b+\G_0\e\times\b)\c
\boldsymbol{S}]\epsilon=\G_{123}\epsilon\,,
\end{equation} 
\begin{equation}
\label{k5b}
-\frac{1}{\l}[(1+\m^2)\G_0-\G_{0123}(\m+\G_{123}\b-\G_0\e\times\b)\c
\boldsymbol{S}]\epsilon=\G_0\epsilon\,,
\end{equation} 
\end{subequations} 
where
\begin{equation}
\label{k6}
\boldsymbol{S}=\n\,\G_9-\e\,\G_0+\m\,\G_{123}\,.
\end{equation} 
The pair  of equations (\ref{k5}) is equivalent to
\begin{subequations}
\label{k7}
\begin{equation}
\label{k7a}
\l=-(1+\m^2),
\end{equation} 
\begin{equation}
\label{k7b}
\boldsymbol{S}\,\epsilon=0.
\end{equation} 
\end{subequations}
These two conditions (\ref{k7a})
 and (\ref{k7b}) are compatible and give rise to 
the preservation of half supersymmetry. Indeed, from Eqs. (\ref{k7b}) and 
(\ref{sol1}) it follows that
\begin{equation}
\label{k8}
\n=G\frac{\bf r}{r^3}\,,
\end{equation}
where $G$ is a constant. Equation (\ref{k7b}) thus assumes the form
\begin{equation}
\label{sol6}
{\hat M}\epsilon=0\,,
\end{equation}
where
\begin{equation}
\label{sol7}
{\hat M}=G\,\mathbb{1}+e_0\,\G^{09}-b_0\,\G^{1239}\,.
\end{equation}
The matrix ${\hat M}$ satisfies the relation
\begin{equation}
\label{sol8}
{\hat M}^2=(e_0^2+b_0^2-G^2)\mathbb{1}+2G{\hat M}\,,
\end{equation}
therefore its eigenvalues must satify the same relation (\ref{sol8}). Thus 
Eq. (\ref{sol6}) has nontrivial solutions only when
\begin{equation}
\label{sol9}
G=\pm\sqrt{e_0^2+b_0^2}\,.
\end{equation}
Owing to  condition (\ref{sol9}) the eigenvalues of the matrix ${\hat M}$ are 
either $0$ or $2G$. On the other hand the trace of the matrix ${\hat M}$ is   
$32G$, consequently it has sixteen $0$ and sixteen $2G$ eigenvalues. Eqs. 
(\ref{sol1}, \ref{k8}, \ref{sol9}) together also change to the identity the 
condition (\ref{k7a}). As a consequence the solution preserves half of the 
supersymmetries.

\subsection{The $\boldsymbol{(p,q)}$ string}

\noindent
The supersymmetric configuration  (\ref{sol1}, \ref{k8}, \ref{sol9}) simplifies 
the expressions for the conjugate momentum $\P$, magnetic intensity $\h$  and 
conjugate scalar field $\v$ resulting in
\begin{equation}
\label{linfiel}
\P=\e\,,\quad\h=\m\,,\quad\v=\n\,.
\end{equation}
Equations (\ref{linfiel}) manifest that the supersymmetry requirements imposed 
on solution are:\\ 
i) the electromagnetic field must be a solution of the linear 
electrodynamic equations, \\
ii)  the scalar field must be a harmonic function,\\ 
iii) the scalar charge $G$ expressed in terms of electric and magnetic charges 
has an $SL(2,\mathbb{Z})$ symmetry. \\
These prescriptions can be used to construct 
multiple $(p,q)$ strings ending at 
arbitrary locations on a brane. It is clear
 that the solution also satisfies 
the Euler--Lagrange equations (\ref{eil-lagr}) and
the  Bianchi identities (\ref{bianc}), 
and is a BPS state.

Charge quantisation \cite{evs1}
\begin{equation}
\label{quant}
\int\star\left(\frac{\delta I}{\,\delta F}\right)=p\,,\quad
\frac{1}{2\pi}\int F=q\,,
\end{equation}
where $\star$ means Hodge duality operation,
connects the constants $e_0$ and $b_0$ with the 
electric $p$ and magnetic $q$ charges  respectively,
\begin{equation}
\label{charg}
e_0=\pi g_sp\,,\quad b_0=\pi q\,.
\end{equation}
The above relations allow to 
express the energy of the configuration in terms of charges. The Hamiltonian 
density ${\cal H}$ of the static solution is equal to
 \begin{equation}
\label{h}
{\cal H}=1+\n^2\,.
\end{equation}
This gives for the energy ${\cal E}$ 
\begin{equation}
\label{e}
{\cal E}=T_3\int dV+\sqrt{(p\,T_F)^2+(q\,T_D)^2}\int d\,\Phi(r)\,,
\end{equation}
where $T_F$ and $T_D$ are Fundamental and Dirichlet string tensions, 
respectively. The first term is the energy of the flat brane with no 
background. The second term has a
natural interpretation  as a  $(p,q)$ string 
attached to the brane and running off to infinity. The total energy is the sum 
of these two terms which is the typical feature of BPS states.

\section{Orthogonal perturbations}

We obtain the equation of small fluctuations $\eta$ along an
additional spatial direction perpendicular to the brane
and string by evaluating the appropriately extended 
Born--Infeld Lagrangian at the BPS background.  The Lagrangian
density then becomes
\begin{equation}
{\cal L}=-\sqrt{(1+\m_c^2)^2+(1+\m_c^2)[({\d}\eta)^2-(1+{\d}{\Phi_c}^2)
(\partial_0\eta)^2]}
\label{4.1}
\end{equation}
with
\begin{equation}
{\d }\Phi_c=\pm G\frac{{\bf r}}{r^3}, \;\;\;
G=\sqrt{(\pi g_sp)^2+(\pi q)^2}.
\label{4.2}
\end{equation}
The $SL(2,{\mathbb Z})$-invariance of $G$, esssentially the tension
of the $(p,q)$-string, has been shown, for instance, in ref. \cite{pol}.

By expanding ${\cal L}$ and retaining the lowest order terms we obtain the
equation
\begin{equation}
-(1+{\d}{\Phi_c}^2)\frac{d^2\eta}{dt^2}+\triangle\eta=0,
\label{4.3}
\end{equation}
where $\triangle$ is the 3-dimensional Laplacian.  Different from
previous considerations this is the equation for
the case with the full $(p,q)$ string in the background.

In spherical coordinates and with
\begin{equation}
 \eta=e^{i\omega t}r^{1/2}
Y_{lm}(\theta,\phi)\psi(r),\;\; r=G^{1/2}e^z,\;\; 
h^2=G\omega^2,\;\;  a=l+\frac{1}{2},
\label{4.4}
\end{equation}
the equation becomes the modified Mathieu equation
\begin{equation}
\frac{d^2\psi}{dz^2}+[2h^2\cosh\;2z - a^2]\psi=0.
\label{4.5}
\end{equation}
For the present case of the D3-brane in Born--Infeld theory the complete
solution of the equation with derivation of the S-matrix
has been obtained in ref. \cite{ptmz}. Although the  
considerations there utilized expansions for large
$\omega^2$, i.e. $h^2$, the resulting S-matrix expressed in terms
of modified Mathieu functions of the first kind, $M^{(1)}_{\nu}(z,h^2)$,
with Floquet exponent $\nu$, was shown to have the unique
property to be identical with that
obtained for small $h^2$ derived in considerations of the
supergravity background of an extremal D3-brane
in refs. \cite{gubh},\cite{mmlz} except for the difference
in parameters.
Thus the case of the D3-brane is a rare case permitting
 an exact  solution. The explicit and exact S-matrix element
for the lth partial wave is \cite{ptmz}
\begin{equation}
S_l=\frac{\sin\pi\gamma}{\sin\pi(\gamma +\nu)}e^{i\pi(l+1/2)}, 
\;\;\; e^{i\pi\gamma}=\frac{M^{(1)}_{-\nu}(0,h^2)}
{M^{(1)}_{\nu}(0,h^2)}.
\label{4.6}
\end{equation}
The Floquet exponent has to be calculated separately in the
cases of $h^2$ small and $h^2$ large.  Expansions in
ascending powers of $h^2$ have a finite
radius of convergence; expansions in descending powers (valid in 
the complementary domain) are asymptotic.

\subsection{Low energy}

In the low energy, small $h^2$,  case and S-waves
\begin{equation}
\nu=a+\frac{h^4}{4a(1-a^2)}+O(h^8), \;\;\; a=\frac{1}{2}.
\label{4.7}
\end{equation}
In ref. \cite{mmlz} it was shown that
\begin{equation}
e^{i\pi\gamma}=\frac{\nu !(\nu -1)!\sin \pi\nu}
{\pi(h/2)^{2\nu}}[1+\frac{4\nu}{(\nu^2-1)^2}(\frac{h}{2})^4+\cdots]
\label{4.8}
\end{equation}
implying  here
\begin{equation}
e^{i\pi\gamma}\simeq \frac{1}{h}[1+O(h^4)],
\label{4.9}
\end{equation}
i.e. $\gamma $ imaginary.
We define the amplitudes of the incident wave, and  reflected and
transmitted waves respectively as
\begin{equation}
A_i=2i\sin\pi(\gamma +\nu), \;\; A_r=2i\sin\pi\gamma, \;\;
A_t=2i\sin\pi\nu.
\label{4.10}
\end{equation}
The transmitted wave at large negative $z$ represents
particles falling into the D3-brane.
In the present case we obttain
\begin{equation}
\bigg|\frac{A_r}{A_i}\bigg|=|\tan\pi\gamma|= 1-O(h^2), \;\;\;
\bigg|\frac{A_t}{A_i}\bigg|=\frac{1}{|\cos\pi\gamma|}=2h-O(h^3).
\label{4.11}
\end{equation}
Thus in the zero energy limit, i.e. $h\rightarrow 0$, one has
total reflection.   
The corresponding fluctuation transverse to brane and string is
\begin{eqnarray}
\eta\sim r^{1/2}\psi&=&r^{1/2}\sinh(z/2)+O(h^2)  \nonumber\\
&=&r^{1/2}\bigg[\sqrt{\frac{r}{G^{1/2}}}-
\sqrt{\frac{G^{1/2}}{r}}\bigg] +O(h^2) \nonumber\\
&\sim &\frac{1}{2{G^{1/4}}}(r-G^{1/2})
+O(h^2)
\label{4.12}
\end{eqnarray}
in lowest order of $h^2$ and is seen to satisfy 
in the case of the fundamental open string and in the  weak-coupling 
limit, in which the dynamics can be studied using
string perturbation theory, and $h^2\sim g_s\omega^2$,  the Dirichlet 
 boundary condition
$\eta =0$ at $r=G^{1/2}\rightarrow 0$ with $g_s\rightarrow 0$.
In this case the frequency $\omega$ can be large compared
with the string mass scale, i.e. $\omega\leq m_s/\sqrt{g_s}$. 

\subsection{High energy}

In the high energy, large $h^2$ case, the Floquet exponent is a
complicated function of $h$ as shown in ref. \cite{ptmz}. However,
we can obtain the limits from formulae derived there.  Thus for large
$h$ one obtains  with $a^2 \simeq -2h^2+2hq $ the large-$h^2$ expression 
\begin{equation}
|S_l|\simeq \bigg|\frac{A_r}{A_i}\bigg|\simeq\frac{{2\pi}(16h)^q}
{e^{8h}\{\Gamma(\frac{q+1}{2})\}^2}\rightarrow 0
\label{14.13}
\end{equation}
and
\begin{equation}\bigg|\frac{A_t}{A_i}\bigg|\simeq|e^{-i\pi(q+1)/2}|=1.
\label{4.14}
\end{equation}
Hence in the high energy limit the S-matrix yields total
absorption. These results confirm the expectations of ref. \cite{cm}
obtained there on the basis of a simplified model.

Finally we observe that the study of the convergent small
$h^2$ expansions shows that these are valid within a radius of
convergence of some number (depending on $a$)  $\times\; h^2$. 
Beyond this the asymptotic expansions take over.  Thus there is
a critical value of $h^2\sim {\rm some \;\, number}$ determining
a critical energy $\omega_{\rm crit}$, i.e.
$$
\omega_{\rm crit}\sim \frac{1}{\sqrt{G}}
=\frac{1}{[(\pi g_sp)^2+(\pi q)^2]^{1/4}}.
$$
In the case of the pure fundamental or p-string, $\omega_{\rm crit}\sim
1/\sqrt{g_s}$ which becomes large in the weak-coupling limit
$g_s\rightarrow 0$. 

Considering the fluctuation $\eta$,  one obtains as the leading
factor
\begin{equation}
\eta \simeq  r^{1/2}exp[\pm\sqrt{2h^2\cosh 2z+2h^2}],
(1+O(1/h^2))
\label{4.15}
\end{equation}
which again reflects the $z\rightarrow -z$ inversion symmetry.

\section{Conclusion}

\noindent

In the above we  considered the
 D3-brane intersecting with the $(p,q)$ string. T-duality 
permits the  construction of  the Dp-brane
 intersecting with the electric flux carrying the 
D(p-2)-brane. The  method applied above  does not allow
 to find an analytic expression 
for the S-matrix in this case. To understand 
this we first note  that the potential acts 
as a sink. The waves can escape to infinity
 or to the origin. In the case of the 
selfdual D3-brane these interior and outer regions are equivalent to the 
effect that Eq. (\ref{4.5}) is invariant under the interchange 
$r\leftrightarrow 1/r$. This symmetry  played
 an important role in deriving 
the  analytic expression and  disappears in considering higher  or lower 
dimensional branes. However,  the following simplification
is available: one 
can arrange the electric and magnetic fluxes in the
 attached D(p-2)-brane in such a 
way that the brane becomes tensionless. This will be
 a higher  dimensional supertube
 and its energy is an additive quantity--the sum
 of the electric and magnetic 
charges in appropriate units. Comparing the potential in 
Eq.(\ref{4.3}) and the Hamiltonian of the static
 solution in (\ref{h}) one observes 
that these  coincide. We 
have not checked whether such a  relation holds for other branes also,
 but certainly the supertube (or its higher
 dimensional analog) in the  background has to 
give a simplified potential. 
\vspace{0.6cm}

{\bf Acknowledgments:}
\vspace{0.3cm}
\newline
The  work of S.T. was supported by the Deutsche 
Forschungsgemeinschaft (DFG).
D.K.P. acknowledges support by the Korea Research 
Foundation under Grant (KRF-2003-015-C00109).

\end{document}